\documentclass[preprint,12pt]{elsarticle}

\usepackage{graphicx}
       
\usepackage{amssymb}

\def\kms{$\rm km\, s^{-1}$}
\def\cm3{$\rm cm^{-3}$}
\def\Ts{$\rm T_{*}$~}
\def\Vs{$\rm V_{s}$~}
\def\n0{$\rm n_{0}$}
\def\B0{$\rm B_{0}$}

\def\erg{$\rm erg\, cm^{-2}\, s^{-1}$}

\def\L12{L$_{12\mu m}$~}
\def\F12{F$_{12\mu m}$~}

\def\Hb{H$\beta$~}
\def\Ha{H$\alpha$~}
\def\Ly{Ly$\alpha$~}
\def\ff{{\it ff}~}

\journal{New Astronomy}

\begin{document}

\begin{frontmatter}



\title{The symbiotic system Z Andromedae :\\
 a  spectral analysis of the anomalous 1984-1986 outburst}


\author{M. Contini}

\address{
School of Physics and Astronomy, Tel-Aviv University, Tel-Aviv, 69978 Israel
}

\begin{abstract}

The visual magnitude profile of the symbiotic system Z And during the 1984-1986 activity period
 appears double peaked and the flux intensity is low compared to outbursts in other epochs.
The detailed modeling of the observed spectra, 
accounting for the  shells ejected by the red giant star, 
 shows that the outburst is intrinsically single but distorted by the
collision at different phases of the white dwarf
 wind with two   close shells. 

\end{abstract}


\end{frontmatter}

\section{Introduction}

Symbiotic  systems (SS) are generally  composed by  a white dwarf (WD), a red giant (RG) star,
 and by  circumstellar and  circumbinary  nebulae. 
 Gas and dust radiation from the nebulae appear throughout
 a large frequency range, from radio to X-rays.

Z Andromedae (Z And) binary  system consists of a  cool giant
of spectral type M4.5 and  a hot compact component with a temperature of $\sim$ 1.5 10$^5$ K. 
Many  periods of activity have been  reported during more than 100 years, in
1915, 1939, 1960, 1984, and 2000 as
  nonuniform eruptions of classical outbursts.
Sokolowski et al. (2006)  claim that
  the {\it outbursts sometimes come in pairs (as in 1984
and 1986)}, or in a series of eruptions with decreasing maximum
brightness and different shapes, separated in time by periods slightly shorter  than the orbital one
 (Kenyon \& Webbink 1984).

The main  characteristics of the 1984-1986 event are
 the two  maxima which  appear   in the  
profile  of the visual magnitude  (Fern\'{a}ndez-Castro et al. 1995, hereafter FC95,
 fig. 1) leading to the ambiguous interpretation of the outburst as a  double  or a disturbed  single  one.
FC95 suggested, on the basis of the 
 line ratio analysis,
 that a drop in the WD radiation flux could  provoke the observed trend of the light curve 
in 1986.
  They  claim that a shell of material was ejected during each of
the {\it two outbursts} at 1984 and 1986.

The  brightness  of the 1984-1986 event    was  exceptionally low 
(Leibowitz \& Formiggini 2008, fig. 1). Nevertheless,  it was monitored
at very close dates by the spectral observations of FC95. 
The  spectra are rich enough in number of lines
to  allow a detailed modeling of Z And physical conditions.

In this paper we  revisit Z And  system
focusing on the  physical conditions in the  nebulae  and their fluctuations
in the 1984-1986  years. 
Our aim is to reveal  unpredicted  episodes
by the detailed modeling of the spectra.

In Z And,
the stellar wind  of the cool giant has a velocity 
of $\sim$ 25 \kms (Sequist et al. 1984). 
 Bisikalo et al (2006) claim that
 varying the wind velocity from 25 \kms to 30 \kms changes the accretion regime
from disk to wind accretion.
This is accompanied by a jump in the accretion rate, increasing the hydrogen
burning rate. An optical thick wind forms from the WD.
This wind   
is   revealed e.g. by  the observations of  UV and optical spectra
 during the 2000-2002 outburst 
(e.g. Sokoloski et al 2006, Tomov et al 2008).

The shocks  which derive from     WD and RG wind collision
yield an increase in the luminosity of the system.
This hypothesis is strengthened by FC95 who did not find evidence for an accretion disk
at the 1984-1986 epoch  but suggested  collision episodes.

Collision of the winds  (Girard \& Willson 1987, 
Bisikalo et al 2006, Angeloni et al 2010, Contini \& Angeloni 2011)
 leads to two  
shock fronts  between the stars : the strong one, dominating the spectrum,  propagates
in reverse towards
the WD,  while the weakest one propagates towards the RG.
Moreover, a shock front  expands out of the system throughout the circumbinary
medium.

We  calculate the line spectra in Sect. 2 in the frame of the wind collision model,
 adopting the  code SUMA\footnote{//wise-obs.tau.ac.il/$\sim$marcel/suma/index.htm},
which   accounts  consistently for   shocks 
 and   photoionization. 
The input parameters  are: the  shock velocity \Vs, the   preshock density \n0,
the preshock magnetic field \B0, the colour  temperature of the hot star \Ts,
the ionization parameter $U$  relative to the black body (BB)  flux
 reaching the nebula.
The geometrical thickness of the emitting nebula $D$,
the dust-to-gas ratio $d/g$, and the  abundances of He, C, N, O, Ne, Mg, Si, S, A, Fe
relative to H are also accounted for.
\B0=10$^{-3}$ gauss is adopted.

The observations  at different epochs of  the continuum spectral energy distribution (SED)  are reproduced
  in Sect. 3  by  
 models  constrained by the fit of the line ratios.   
The adjustment factors determine 
the  distance of the  nebulae downstream of shock fronts from the
system  center,  providing
 a detailed picture of the  Z And components.
Discussion and concluding remarks follow in Sect. 4.

\section{Line spectra}

The  trend  of the spectral observations  between 1978 and 1993  (FC95, figs. 4,5)
gives  a first  hint about  the 1984-1986  outburst  in the frame of a longer  activity period.
We notice that :

\noindent
1)  the 1984-1986 outburst lies upon an event
which   developed about at 1979 and ended at  1988. If this long event depends on the WD
activity,  then the temperature of the WD  will not show
dramatic changes in the 1984-1986 period. 

\noindent
2)  The OI 1305 resonance line is
  most probably blended  with  an upper ionization level line because
its behaviour is similar to that of  high level lines
The OI line  is most probably  emitted from a different nebula
with conditions  close to   those of the ISM.

\noindent
3) the systematic decrease and increase  of the CII 1336 line, whose trend is opposite 
to that of the high ionization lines, indicates that the CII minimum is strongly correlated
with the trend of the physical parameters, e.g. 
the ionization parameter, and the shock velocity.
The role of the WD temperature  is  less prominent.

\subsection{The UV-optical line spectrum at July 11 1986}

FC95  in their table 3 report  an optical spectrum in July 1986,  observed at the same 
epoch  as the UV spectrum presented in their table 2B, row 34.

We start by modeling  the UV-optical combined spectrum observed at 11 July 1986 which is constrained
by  a relatively large number of lines from various ionization levels. This
reveals the physical conditions in the emitting nebulae
at that time.  Such conditions will be adopted  as a first  guess in the modeling  of the UV 
spectra in the next epochs.

\begin{table*}
 \caption{The UV and optical  line ratios to \Hb (July 11 1986)}
\begin{tabular}{lllllllll}\\ \hline  \hline
line         &  obs  &m$_{34*}$ &m$_{34}$ &&  line         &  obs  &m$_{34*}$ &m$_{34}$  \\
\hline
\ NV 1240    & 3.7 & 55.&4.4 &&   MgII 2800 &+    &-& -  \\
\ OI 1306    & 0.74&0.77& 0.01&&  [NeV] 3425 & 0.2 &0.3&0.15     \\
\ CII 1336   & -   &-&-&& [NeIII] 3869&0.1&0.25&0.04    \\
\ OV  1370   &  0.18&95.& 0.3 &&  H$\gamma$ 4360& 0.31 &0.38    &0.4     \\
\ SiIV] 1400 & 1.8  &42.& 1.6  && [OIII] 4363 & 0.04 &0.6&0.03\\
\ NIV]  1486 & 0.9  &6.3& 1.1  && HeI 4471     &0.087&0.005& 0.024 \\
\ CIV  1550  & +    &-&- && HeII 4686  & 0.6  &0.6& 0.67\\
\ HeII 1640  & +    &-&- && HeII 4686  & 0.6  &0.6& 0.67\\
\ OIII] 1643 & 1.0  &29.&0.33 && \Hb  4861  & 1.   &1.& 1.   \\
\ NIII] 1750 & 0.4 &30.&0.2  && HeI 5876   &0.157 &0.135&0.08       \\
\ SiIII] 1890 &0.5 &28.&0.7  && [FeVII] 6087&0.11 &0.16&0.04    \\
\ CIII] 1910 &0.8   &48.&1.1 &&    H$\alpha$   & 6.2&5.6& 3.1      \\
\hline
\end{tabular}
\end{table*}

The  optical - UV spectrum   at July 11 1986
is presented in Table 1.  The lines are normalized to \Hb=1.

Table 1 shows that the [OIII] 4363 line is not negligible, while the
[OIII] 5007+4959 lines which  are generally the strongest ones   were not observed.
This indicates   a very high pre-shock density ($\sim$ 10$^{8-9}$ \cm3)
 characteristic of  nebulae  in SS (Angeloni et al. 2010).

The   line intensities were reddening corrected  adopting E(B-V)=0.35.
Before correction \Ha/\Hb =9.3, after correction  \Ha/\Hb = 6.2  is still  high compared to $\sim$ 3  
which results in the optically thin case (case A, Osterbrock 1989). 
Only in case A the lines with a relatively low critical  density for collisional deexcitation
(e.g. [OIII] 4363)  can be observed.
A higher E(B-V) would  result in abnormally high far UV line fluxes. Moreover, the observations in the IR
do not suggest a large obscuration.

High  \Ha/\Hb line ratios are typical of symbiotic stars. They
 could  be due to radiative transfer effects of HI, to collision excitation of H lines, and, 
more specifically for symbiotic stars, to the effect of the viewing angle of the system (Contini 2003).
In the following, we will consider the UV spectra  independently from the \Ha/\Hb question.

Modeling  the optical-UV spectrum, we
  have   first  fitted the  ratios of the nitrogen lines, because  N appears in three
ionization levels (III, IV, and V), then we have considered the [NeV] and [NeIII] lines.
 The HeII lines 
depend strongly on the photoionization flux. The OV, SiIV], and SIV lines are   blended
with  the OIV] multiplet.
The OI 1305 line is most probably blended with the SiII 1263,1308 doublet.

Model m$_{34*}$ (Table 1) shows \Ha/\Hb $>$ 5 and reproduces most 
of the observed optical lines, except [OIII] 4363/\Hb which is overpredicted
  by a factor of 10. 
The spectrum  is obtained by  interrupting the downstream region at a distance of 5 10$^8$ cm
from the shock front. The input  parameters are \Ts=1.2 10$^5$ K, $U$=0.1, \Vs=80 \kms, \n0= 10$^8$ \cm3,
\B0=10$^{-3}$ gauss, and solar abundances. The density throughout the
whole emitting region downstream  is $>$ 10$^9$ \cm3.

Although  the [NeV]/[NeIII] line ratio is acceptable, most of the UV line ratios to \Hb are overpredicted 
by a factor $>$ 10.  Even  assuming that the lines in the UV are emitted from a region different
from  that  emitting the optical spectra, recall that all  the spectral contributions must be 
summed up, adopting the same relative
weights (see Contini et al 2009b) for  UV and optical lines. This can produce  less fitting results.
A different reddening correction
cannot help in this case, because it would  overestimate  the [OIII] 4363 line in the optical range.

Recent results  obtained by   modeling  the   AG Dra spectra (Contini \& Angeloni 2011)   imply 
 collision of the WD wind with   the  shells ejected  from the red giant.   
The shells contain grain and  molecular dust  leading to the 
underabundance of C, N, O,  Si and Mg in the  downstream gas because these elements  are trapped into 
dust grains and  molecules. Ne  is not  adsorbed in  grains due to its
atomic structure. Fe is also present in the gaseous phase.

Sputtering is not effective at  velocities as low as those   calculated from the FWHM of the line profiles.
Fragmentation of matter at the shock front  produces dense  clumps of gas and dust.
Adopting C, N, O, and Si abundances lower  than solar by a factor $>$ 10 
relative to H, as for AG Dra, (Contini \& Angeloni 2011) would improve the  fit  of the Z And spectrum. 
However,   there is no other evidence  to justify this assumption. 
Errors in the   observations and
uncertainties   of  the atomic coefficients adopted in  the calculations cannot  yield such discrepancies.
So we will start by modeling the UV line ratios
with relative abundances close to the solar ones
and we  will  check  the validity of this hypothesis  by 
the  spectra observed  in the UV. 

 The model  best fitting  most of the UV and optical observations at 
July 1986 (m$_{34}$) is presented in Table 2,  last row.    

\subsection{UV spectra}  

The UV spectra  observed by FC95  and presented in their table 2B  were reddening corrected and are 
shown  in Table 3. They correspond to the following dates : JD 2440000+
4719  :   24/Apr/81 ;  +5797:    06/Apr/84  ; +5864  :    12/Jun/84 ;   +5948  :  04/Sept/84 ;
 +6233  :   16/Jan/85 ;    +6283  :   05/Aug/85 ;  +6353  :    14/Oct/85 ;  +6374 :  04/Nov/85 ;
 +6402  :   02/Dec/85 ;  +6436 :  05/Jan/86 ;  +6462 :    31/Jan/86 ;   +6623 :    11/Jul/86.

We  selected the spectra observed between Apr 24 1981 and Jul 11 1986 with the maximum number of  lines, i.e.
with the least number of saturated (+) and/or not detected fluxes (-). The spectrum observed at
Nov 19 1980   is poor in number of lines,
so we  adopt   April 1981 as the start of the first peak.

The three nitrogen  lines are relatively strong and
the NIV] line  is present in  all the significant spectra observed in  the different epochs.
Therefore,
we   normalise the line ratios to the NIV] 1486 multiplet summed terms,   avoiding the element abundance
problem at least  for  N. 
CIV 1550 which is generally a strong line, is saturated in many spectra.  Both C and Si can be locked into grains
so   C and Si lines are less adapted to lead the modeling process.
We wonder why the OIV] 1394-1407 multiplet was not observed and we suggest that the OIV] lines are blended 
with OV 1370 and with SiIV 1400. The S lines are not reported by FC95. 

The observed line ratios  to NIV] 1486 are compared with  calculated  line ratios    
 in Table 3.
The  row containing the data at each epoch  is followed  by that showing the spectrum 
calculated by   the  model  indicated in the  first column.
We reproduce the line ratios within a factor of $\sim$ 2.
The fit of the line spectra and of the continuum SED  at different days  are cross-checked (Sect. 3).

The models are  presented in Table 2. They were all calculated adopting solar abundances
(He/H=0.1, C/H=3.3 10$^{-4}$, N/H=9. 10$^{-5}$, O/H=6.6 10$^{-4}$, Ne/H=10$^{-4}$, 
Mg/H=2.6 10$^{-5}$, Si/H=3.3 10$^{-5}$,
S/H=1.6 10$^{-5}$, Ar/H=6.3 10$^{-6}$, Fe/H=3.2 10$^{-5}$, Nussbaumer et al. 1988),
 except   models m$_{22}$, m$_{23}$, and m$_{30}$, where C/H=2.3 10$^{-4}$.
In the last two  columns of Table 2 the calculated absolute flux of \Hb and   the \Hb/NIV] line ratios  
would provide the  absolute flux values for all the lines.

The spectra show relatively strong lines from   high ionization levels (e.g. NV 1240) as well as strong
neutral lines (e.g. OI 1305) at the same time,
indicating that the lines are emitted by different nebulae, as
it was found for e.g. 
H1-36, AG Dra, and BI Cru  (Angeloni et al 2007, Contini \& Angeloni 2011,  Contini  al. 2009c).
For Z And,  a single-nebula model (m$_{34}$) in July 1986 and in the other epochs  explains most 
of  the  observed spectrum, except the OI 1305 line.
 Table 3  shows that the contribution of a shock dominated nebula (model mm)
characterised by a very low ionization parameter
and high pre-shock density improves drastically the OI 1305/NIV] fit. 
The elements which can be trapped into dust grains,
Si, Mg and C, are depleted from the gaseous phase, while  oxygen is only slightly depleted (Si/H= 3.3 10$^{-6}$
Mg/H=2.610$^{-8}$, C/H= 2. 10$^{-4}$  and  O/H=6. 10$^{-4}$).
This model  also improves the modeling of other line ratios,
such as SiIII] 1890 at Apr 1984, June 1984, Jan 1985, Aug 1985 and Oct 1985 if summed up to 
 models m$_{13}$-m$_{34}$ adopting different weights.

Models m$_13$-m$_{34}$ represent the nebula downstream of the shock front propagating towards the WD,
while  model mm  represents dusty  clumps created by fragmentation of matter  by collision with
RG debris.

\begin{table*}
\caption{The models \label{tab:mods}}
\begin{tabular}{lccccccccc}\\ \hline  \hline
\  model &  \Vs  &  \n0  & $D$  & \Ts  & $U$ & \Hb  & \Hb/NIV] \\
\   &   (\kms)  &   (\cm3) &  (cm) &  (K) &  - &  (\erg) &  -\\ \hline
\ mm &  150         &5.e8         & 8.7e13   &   1.2e5     & 0.0025& 8.7e6  & 2.e3\\
\ m$_{13}$&  120        & 4.e7         & 2.e14    & 1.3e15      & 3.    &4.e5    & 1.   \\ 
\ m$_{15}$&  140        & 3.e7         & 3.6e13   & 1.4e5       & 2.    &7.7e5   & 1.73 \\ 
\ m$_{17}$&  210        & 6.e8         & 3.e9     & 1.4e5       & 0.5   &1.1e4   & 0.7  \\
\ m$_{20}$ & 210        & 6.e8         & 3.4e9    & 1.4e5       & 0.6   &1.1e4   & 0.68\\
\ m$_{22}$ & 300        & 4.e8         & 1.3e10   & 1.7e5       & 0.8   &1.3e4   & 1.3 \\
\ m$_{23}$&  300        & 2.e8         & 1.7e10   & 1.6e5       & 0.4   & 6200.  & 1.15\\
\ m$_{24}$&  150        & 5.e8         & 1.1e9    &$>$1.5e5     & 0.2   & 6738.  & 0.55\\
\ m$_{25}$&  110        & 1.e8         & 1.9e9    & 1.3e5       & 0.01  & 384.   & 0.52\\
\ m$_{26}$&  160        & 6.e7         & 4.4e9    & 1.3e5       & 0.08   &335.   & 0.45\\
\ m$_{28}$&  100        & 5.e7         & 9.e9     & 1.3e5       & 0.08   &130.   & 0.18 \\
\ m$_{30}$&  150        & 8.e7         &6.4e9     & 1.3e5       & 0.2    & 974.  & 0.376 \\
\ m$_{34}$&  100        & 7.e7         & 8.e14    & 1.2e5       & 1.7    &5.9e5  & 1.1   \\ \hline
\end{tabular}
\end{table*}

\begin{table*}
\caption{Comparison of calculated with observed line ratios to NIV]=1 \label{tab:spect}}
\small
\begin{tabular}{clllllllllllll}\\ \hline  \hline
 JD$^1$ &  NV  & OI       &   CII    &   OV    &    SiIV  &   CIV  &    HeII   &    OIII] &   NIII   &  SiIII] &   CIII] &   MgII \\
           &  1240 & 1305     & 1335    &  1370   &  1400    &  1550  & 1640      &  1663    & 1750     & 1890    & 1910    & 2800   \\ \hline
  4719     &  8.4  & 0.69     & -       &   -     &  2       &  9.6   &  7.9      &  1.03    & 0.24     & 0.62    & 0.81    &  -    \\        
m$_{13}$  &  7.7  & -        &  -      &   -     &  1       &   8.7  &  7.1      &  0.34    & 0.2      & 1.4     & 1.8     &  -    &  \\ 
  5797     &   2.75&  0.53    &0.03    & 0.24    & 1.7      &  5.91  &  +        &  1.      & 0.55     & 0.66    &  0.7    &   + \\             
m$_{15}$   & 2.6   &  -       &    0.2 &   0.4   &  0.9     &    12.2&   12.9    &    0.7   &     0.4  &   2.3   &   3.    &  -     \\    
  5864     &  4.77 &  0.97    &0.03    &  0.85   & 0.9      &   +    &    +      &    +     & 0.5      &  0.59   &  +      &    +  \\             
m$_{17}$   &  4.7  &  0.05    &0.24    &  0.43   & 3        &  -     &   5.3     &   1.2   & 0.5       & 6.3     &  -       &   -  \\ 
  5948     &   4.0 &  0.98    &0.02   &  -      & 2.3      &  +    &  5.3      &   1.     & 0.4       & 0.54    & 0.59    & 0.09   \\             
m$_{20}$   &   4.6 &  0.04    & 0.16   & -       &    3.    &  -    &   5.2     &      1.  &      0.4  &    0.54  &  0.54  &    0.13 \\
 6233      &   8.8 &  0.17    &  0.04  &   0.4   &  3.6     & +     &   8.28    &    0.97 &   0.33    &    0.43&     0.7 &   +     \\    
m$_{22}$  &   7.4 &  0.1     &    0.3 &   0.7   &   4.     &  -   &    9.8     &     1.58 &   0.6     &   10.   &      0.8&   -     \\
  6283     & 5.87  &  1.      &   -    &  0.58   &   2.56   & 9.9  &   7.17     &    1.03 &   0.38     &   0.59  &  0.95   &     -   \\        
m$_{23}$    &  5.3  &  0.1     &  -     &  0.5    & 2.2      &   1.6 &       7.7 &     2.7 &      1.    &     10. &     1.7 &   -      \\
 6353      &  3.9  &  0.52    & 0.07   &  0.18   & +        &  11.5 &   4.56    &    1.3  &    +       &0.3      & 1.4     &   +     \\       
m$_{24}$   &  3.   &  -       &    0.1 &  0.3    &      5.2 &   17. &      4.0  &   1.47  &   -        &    5.   &      1.2 &    -    \\
 6374      & 1.17  &1.1       &0.24      &    -    &   4.85   &     14.&   1.23   & 3.57    &  2.54      & 3.0     &  4.96    & 0.5     \\
m$_{25}$  & 1.2   &0.1       &     2.6  &    -    & 4.5      &   14.  &    1.25   &  3.1    &    2.     &   3.6   &   5.2   &    2.4     \\
 6402      &   4.29&2.        &  0.5     &   -     &  5.68    &   11.3 &   1.35    & 3.88    &   2.65    &   4.16  &   5.06  &   +      \\ 
m$_{26}$  &   3.8 &0.1       &    0.8   &   -     &  6.5     &  13.3  &    1.1   &   2.8    &   1.8     &   3.    &   4.    &    -      \\
 6436      & 0.87  &1.48      &  0.63    &   -     & 4.19    & 7.       &    0.51  & 2.59      &   2.09  & 4.07    &   3.53   &  0.27 \\
m$_{28}$  &   1.   &0.1      &0.2     &   -     &  4.       &   11. &   1.3     &1.2      &       1.  &       2.&     3.2 &     0.12  \\
 6462      & 0.8    &1.51     &   -     &   -     &  4.16     &   5.63 &  1.53    &2.16     &  1.78     &     +   &    +    &   0.14   \\
m$_{30}$   &  1.6  &  -       &    -     &   -     &  4.7      &   10.  &      2.6 &     1.6 &      1.   &     3.5 &    -    &      0.23  \\  
 6623      &   4.03& 0.8      &   -     &  0.2    & 2.26      &   +    &    +     & 1.09    &  0.46     & 0.48   &  0.84   &    +       \\
m$_{34}$  &  4.4  & 0.01     &   -     &  0.3    &  1.6      &   -    &    -     &   0.33   &      0.2 &  0.65   &   1.1   &    -      \\ \hline
 mm   &   4.6 & 1.1      &   5.     &  1.      & 3.       &   10.4 &   0.8    &  2.6    &  0.8     & 0.49    & 1.1     &    3.2   \\ \hline 
\end{tabular}

$^1$ 2440000+

\end{table*}

\begin{figure}
\begin{center}
\includegraphics[width=0.98\textwidth]{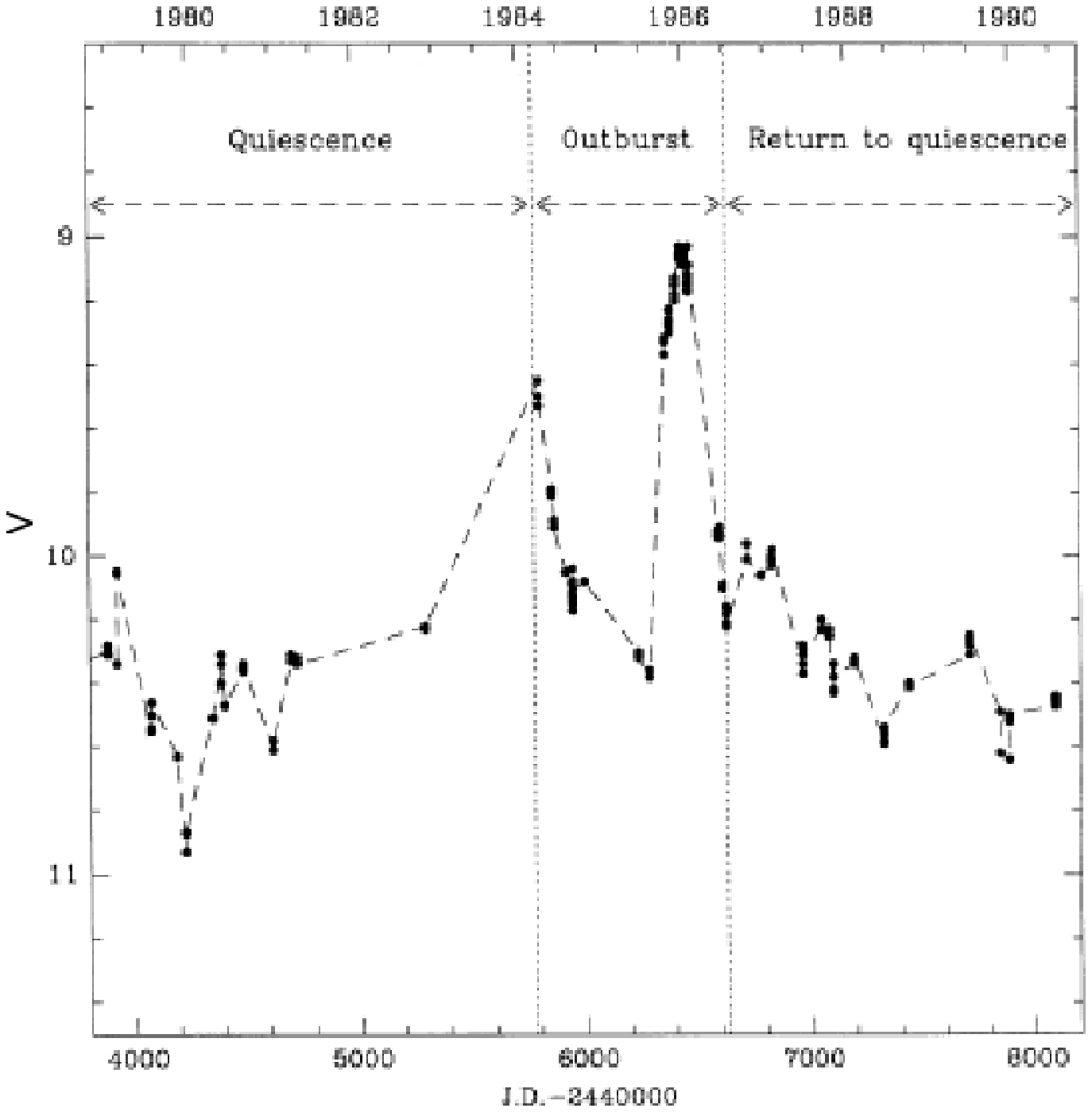}
\caption{Top : the FES  light curve of Z And adapted from FC95
(fig 1).
Dotted black lines define the outburst.}
\includegraphics[width=0.88\textwidth]{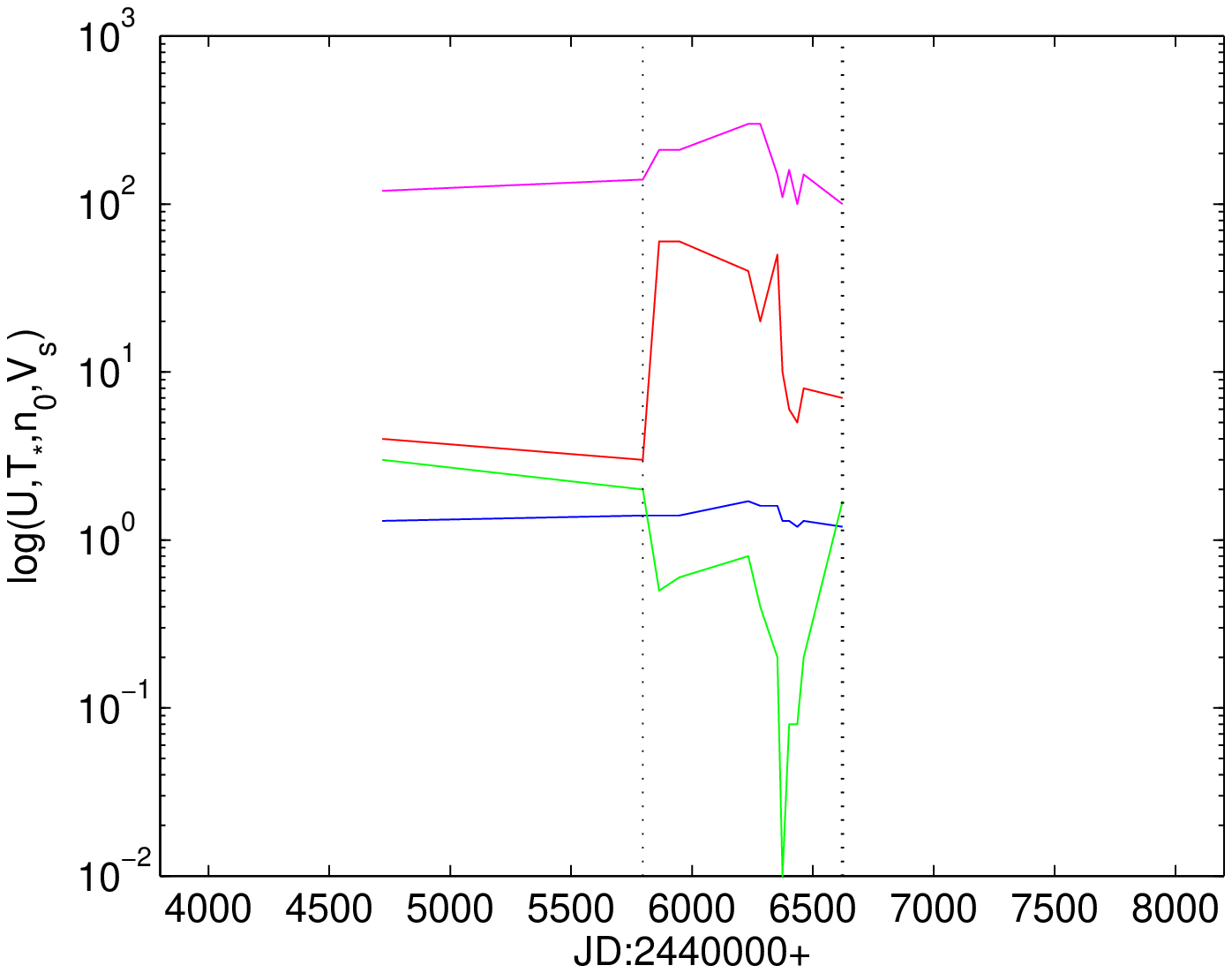}
\caption{
The physical parameters (in logarithm scale) across the outburst calculated
by modeling the line spectra. \Vs : magenta; \n0 (in units of 10$^7$ \cm3) : red; \Ts
(in units of 10$^5$ K) : blue ; $U$ : green.
Dotted black lines define the outburst.}
\end{center}
\end{figure}

\subsection{Results}

The  visual light curve (Fig. 1) covers a frequency band 
accounting for lines and continuum in the optical range.
The strongest  lines are generally HeII 4686, \Hb, [OIII]5007, 4959,  etc., while
the optical spectrum  (FC95) shows relatively low
forbidden  lines. So we consider that the main contributors to the
visual band are HeII 4686 and \Hb, which depend strongly on the WD temperature
and on the ionization parameter.
 \Ts and $U$  affect the spectra in   the same way.
Namely, when they decrease, the high level lines  decrease relatively to the low level ones.

The  increase of \Ts at the epochs of the peaks and the  decrease at minima 
would  indicate that the active phase is characterised by two different  bursts.
On the other hand, if \Ts does not change significantly and only $U$ changes,  
a disturbing  dynamical and/or  morphological event   between the radiation source
and the nebula  is most likely   the source of the  visual curve minimum.

The UV band  accounts for lines from different levels which are variously modified
by \Ts and $U$ fluctuation. Moreover, the shock velocity and the compression downstream
affect the stratification of the ions  and, consequently,
the line ratios.
FC95 show that lines from different ionization levels have different trends
throughout the 1984-1986 time period.
We predict that the  UV light curve throughout the outburst,  would have a  profile 
different  from that reported in Fig. 1.

In Fig. 2  we present the  physical parameters  obtained  by modeling  the spectra 
throughout the 1984-1986 outburst. 
The profiles of the physical conditions  are  rather unexpected if compared
with  the  visual light curve (Fig. 1).
 
The  sudden increase of the density on April 1984, accompanied by the sudden increase of the velocity, 
indicates that collision of the WD wind with a   shell ejected from the red giant  is occurring.
The nebula downstream of the reverse shock is heated and ionized by both the shock and the photoionization
flux from the hot star. In the next days
the density and the shock velocity have opposite trends in agreement with conservation of mass 
at the shock front (n1v1=n0v0, from the Rankine-Hugoniot equations).
 In the meanwhile the temperature of the hot star is slightly
increasing from $\sim$ 120,000 K to a maximum of 170,000 K, corresponding to the outburst maximum,
in agreement with FC95.
 The drop  of $U$ on 1984 toward July 1986  shows
 that the spectra   during that epoch are  dominated by a relatively strong shock, i.e. 
 the  reverse shock, which vanishes at a certain time, depending on the outburst characteristics
 (Chevalier 1982).  

The second density peak at December 1985-January 1986 reveals that  collision  of the WD wind with another
RG shell has occurred.
The  ionization parameter  drop by 
 more than two orders of magnitude after June 1985,   
 accompanied by the simultaneous variation   of the dynamical
parameters  \Vs and \n0,    suggests  that the
  shock front has suddenly   shifted farther from the WD.

 The matter downstream of the reverse shock front  
 is  fragmented by Richtmyer-Meshkov and Kelvin-Helmholtz instabilities. 
High density fragments between the stars
 eventually  prevent the black body flux from reaching the nebula.
However, the decline of the BB flux by  fragment obstruction 
could indeed reduce the ionization parameter, but  less  sharply than predicted by the calculations.
So, the whole picture   is  most likely explained by an  unexpected disturbing episode
such as the ejection of a new shell from the RG atmosphere and its collision with the WD wind at a relatively 
large distance from the WD (Sect. 3).

The gas composition at the epochs of collision with the  shells reveals that
the calculated
SiIII]/NIV] line ratios would better reproduce  the data adopting  
Si/H relative abundances lower than solar by a factor of 4 for models m$_{13}$, m$_{15}$ and 
of $\sim$ 10 for models m$_{17}$, m$_{22}$, m$_{23}$, m$_{24}$.
This  indicates that silicate grains are present
after collision of the WD wind with the first RG shell.  On the other hand
 Si is in the gaseous phase with a solar Si/H relative abundance
at  the epochs corresponding to models m$_{26}$, m$_{28}$, m$_{30}$, m$_{34}$.
The grains  had  not enough time to form at collision with the next shell. 
Grain formation time scales  (Gail et al 1984, Scalo \& Slavsky 1980) are as long as $\sim$ 1 year. 
 However, in circumstellar shells the chemical equilibrium 
and consequently grain formation is  disturbed by the UV radiation from the WD and by 
the velocity field (Gail et al. 1987).

\section{The continuum}

In Fig. 3 we  compare the observed continuum SED  with model calculations.

\subsection{The continuum SED}

We refer to the SED of the continuum flux 
presented by FC95 in their fig. 8.
We have   used the data   in the UV and in the radio range presented by  FC95
tables 2A and 4, respectively.
The  dates corresponding to  observations in the radio and those in the other  frequency ranges
 are not exactly coincident, e.g.  we adapted the
radio fluxes observed in October 1986 to data observed in July 1986. 

Each diagram  of Fig. 3   presents the   free-free + free-bound fluxes emitted from the nebula  
at a certain epoch.
They are calculated by the same models (Table 2)  which lead to the best fit of the line spectrum
(Table 3). Moreover, the BB flux from  the RG corresponding to a temperature of 3200 K (FC95),
and the BB flux corresponding to the WD which was derived by modeling the line spectrum
are also shown.
Reprocessed radiation of dust within the nebula calculated consistently with gas emission  is added in Fig. 3 diagrams.  
 This  flux is mostly hidden by the BB flux from the RG  if dust-to-gas ratios of  10$^{-14}$ by number
(4 10$^{-4}$ by mass for silicates) are adopted. Actually, the dust-to-gas ratios
are constrained by the data in the mid-IR.

Fig. 3 shows that the emission from the nebula  appears in a small frequency range
between the optical and   the near-UV ($\sim$  1-3 10$^{15}$ Hz), depending  on the
epoch. There are no data in the soft X-ray range and beyond it. We can predict that bremsstrahlung from the
nebula downstream of shock  fronts with \Vs  $\geq$ 160 \kms would be the X-ray source.
For lower velocities the SED in the soft X-ray range is the summed flux of bremsstrahlung
from the nebulae and BB flux from the WD.
The  contribution to soft-X-rays up to the X-ray domain  of bremsstrahlung from shocked nebulae
was suggested   for the symbiotic system AG Dra by Contini \& Angeloni (2011).

The fit of calculated to observed data is good enough
 to  confirm a strong self-absorption of free-free radiation from the nebulae for frequencies
 $<$10$^{13}$ Hz.

Finally, Fig. 3 diagrams show that the  BB  flux from the RG and from the nebula    dominate  the SED in the visual,
while  the flux from the WD  appear in the UV and soft X-ray range.
This indicates that strong variations in the WD temperature
that generally accompany the outbursts cannot be directly revealed by the visual light curve presented by FC95,
but only indirectly by the spectra emitted from the nebula.

\subsection{Radius of the nebulae}

In  previous works on  SS, e.g. CH Cyg (Contini et al 2009b), AG Dra
(Contini \& Angeloni 2011),  etc.
we  calculated the distance r of the nebulae from the center of the SS
 adjusting
 the continuum fluxes calculated  by the models  at the nebula to those observed
at Earth.  We    defined the adjusting  factor 
 $\eta$  by r$^2$=10$^{-\eta}$ d$^2$ \ff,
where \ff is the filling factor and d the distance to Earth.

Adopting  a distance of Z And  to Earth of 1.12 kpc
(Viotti 1982), the
 best fit to the data of the  bremsstrahlung calculated from the nebulae downstream of the shock front,
gives   r= 1.9 10$^{14}$ cm at April 1981,
r=1.4 10$^{14}$ cm at April 1984, r=9.4 10$^{15}$ cm at December 1985, and r= 2.7  10$^{14}$ cm
at July 1986. The distances were  calculated by a filling factor \ff =1.

Considering that : $F_{\nu}$ (R$_{WD}$/r)$^2$ = $U$ n c (where $F_{\nu}$ is the flux from the WD, 
R$_{WD}$ is the radius of the WD,
n is the density of the gas and c the speed of light), 
the results  suggest that the drop of the ionization parameter $U$
between November and December 1985  most probably   derives   from  the  sudden increase of 
the distance  from the hot source. This is  explained by the ejection of the next  shell  in the RG atmosphere.  

The collision between the wind from the WD and the two shells from the RG
 occurred within less that 2 years, from March 1984 to
November  1985. The shock front has therefore expanded  from $\sim$ 2 10$^{14}$ to  $\sim$9 10$^{15}$ cm,
by  an average velocity of $\sim$ 130 \kms, in rough  agreement with the  observations.

\begin{figure*}
\begin{center}
\includegraphics[width=0.55\textwidth]{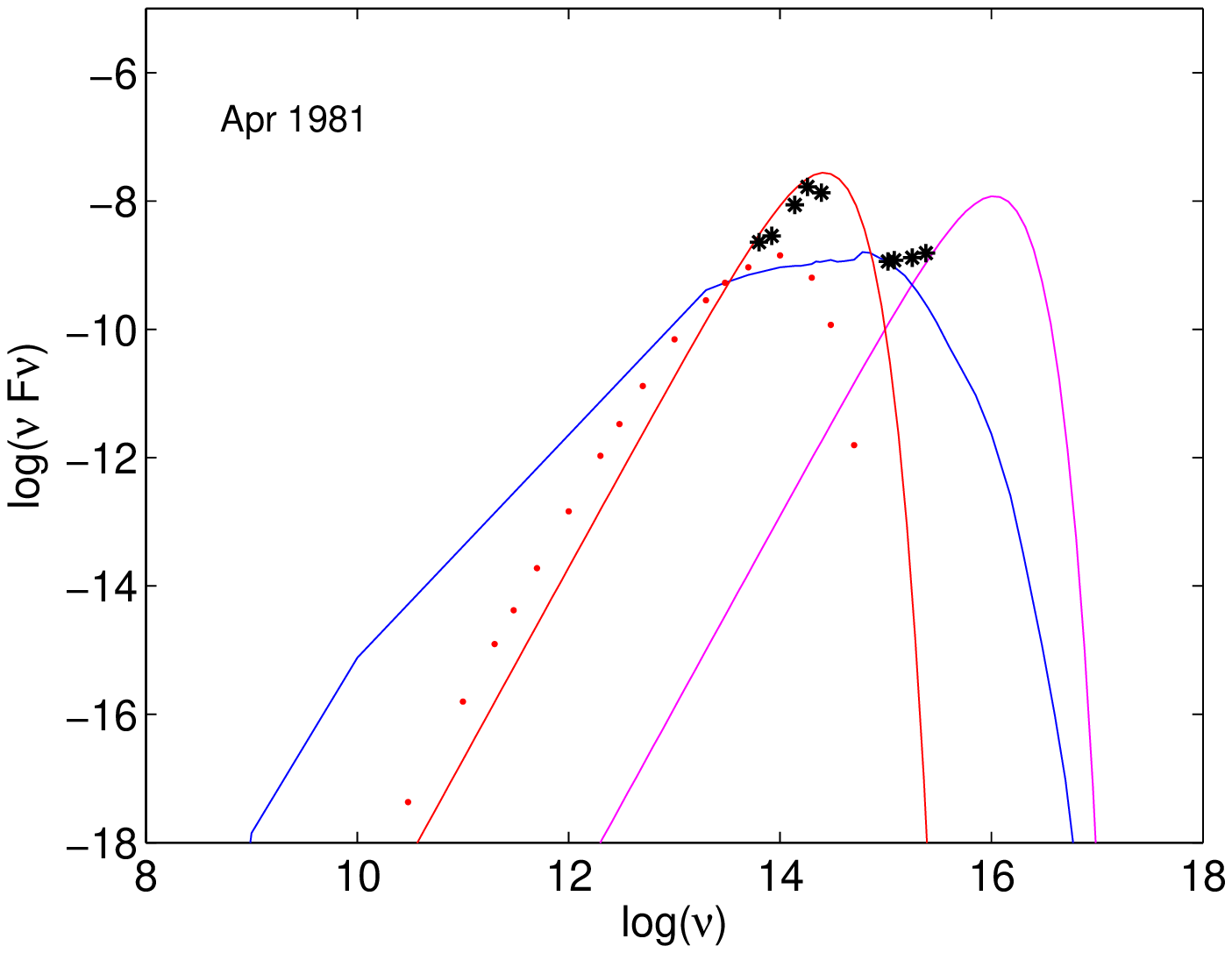}
\includegraphics[width=0.55\textwidth]{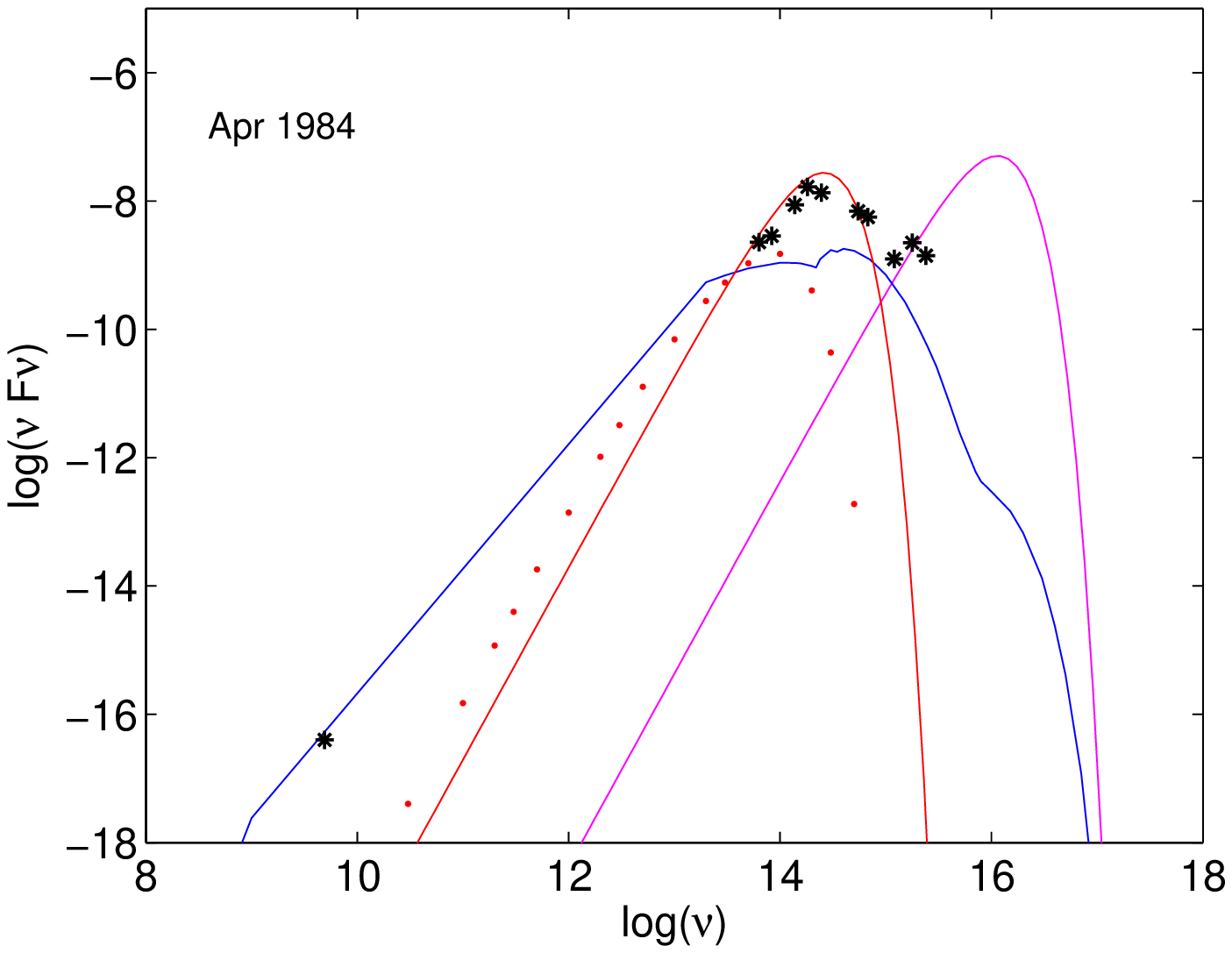}
\includegraphics[width=0.55\textwidth]{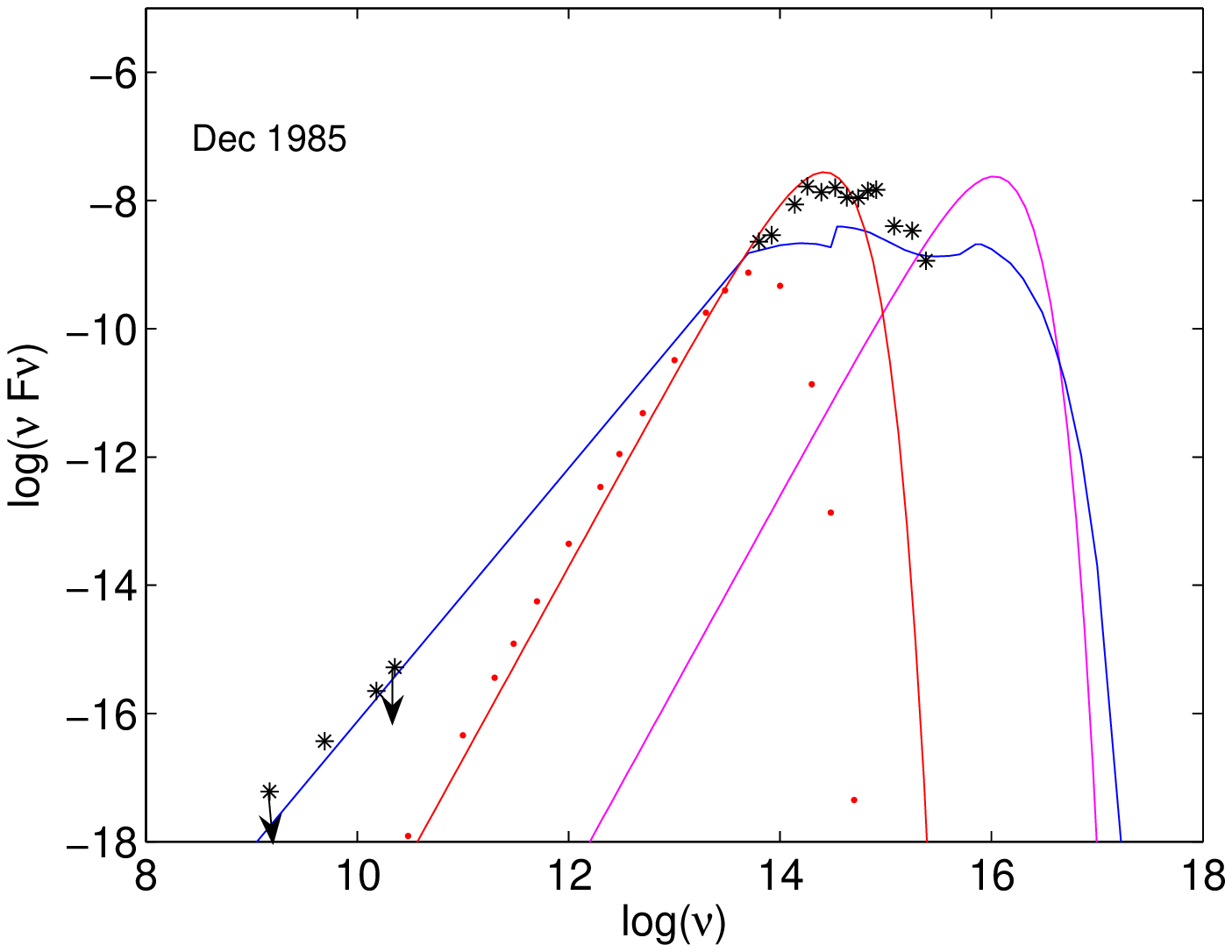}
\includegraphics[width=0.55\textwidth]{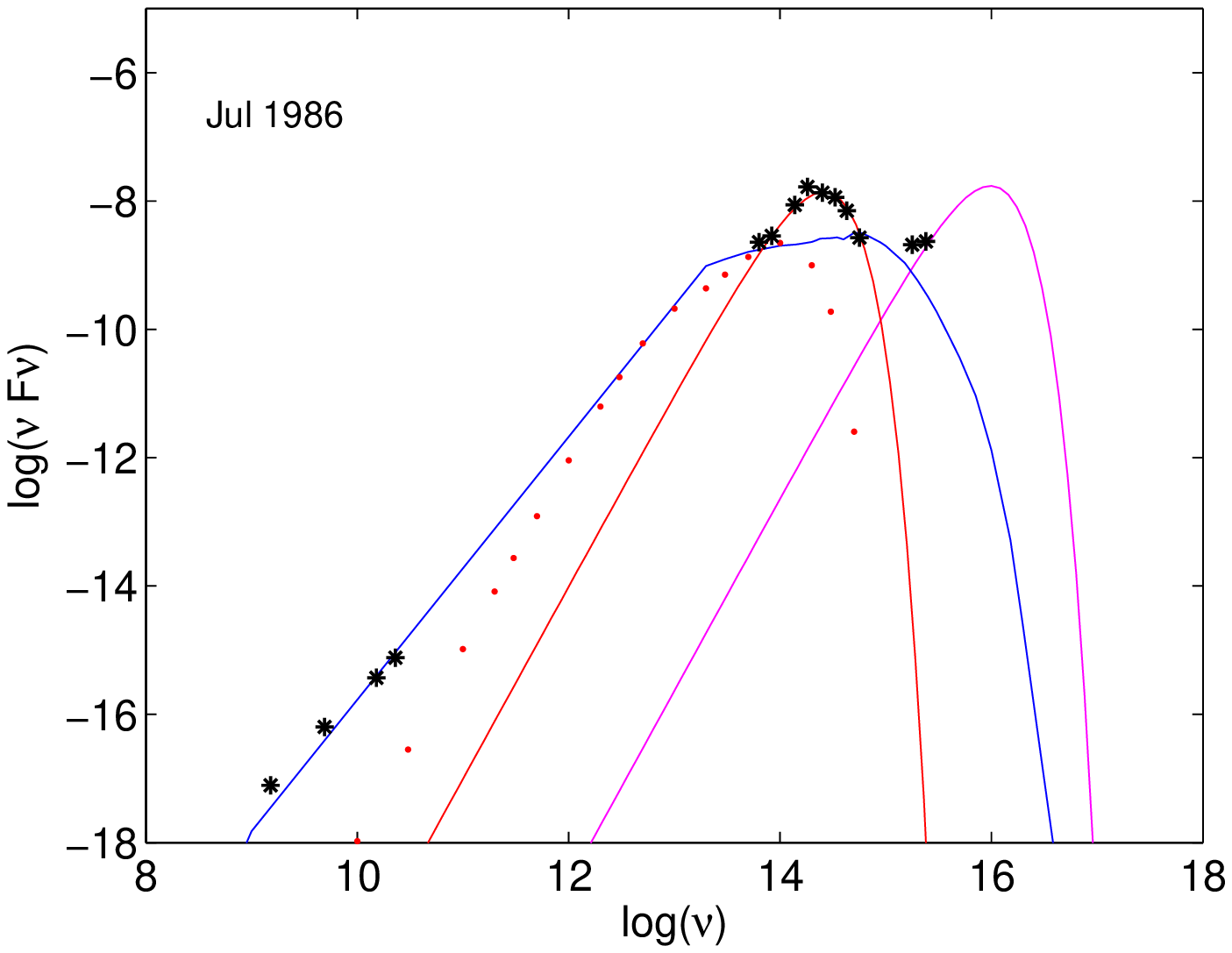}
\caption{ The continuum SED in the different epochs. 
Red giant bb flux (T= 3200 K) : red line; 
 BB flux from the WD : magenta;  bremsstrahlung from the nebulae : blue;
  infrared dust reradiation flux : red dots; data : black asterisks.
 }
\end{center}
 \end{figure*}

\section{Discussion and concluding remarks}

The 1984-1986  outburst  of Z And is revisited by a detailed modeling of the spectra.
Comparing   the visual light curve of Z And during the 1984-1986  outburst with those in other epochs 
from  1895 up  to 2007 September,
 the 1984-1986  event is remarkable 
  for  its double peaked structure and for its low brightness.

We explain the burst double structure in the light of red giant pulsation, 
by the collision of the WD wind with two  ejected shells.
The
 minimum of the light curve between the two  peaks is accompanied by the dip 
of the ionization parameter   which  is constrained
 by the detailed modeling the UV line spectra at November-December 1985.
The sharp drop of  $U$ is due to the sudden increase of the distance between the WD and the
downstream nebula reached by the WD black body flux, namely,
 the collision of the WD wind with  a new  shell  in the RG atmosphere.
Since the  shell  was recently formed at the time of observations, the grains  did not  have enough time to develop. 
This is revealed by the solar relative abundance of Si/H.

We suggest that the outburst did not attain its maximum luminosity because the collision of the 
wind network  was  distorted by  the  oncoming of the next shell.

The small peaks in the  light
curve before and after the 1984-1986 burst are  most probably due to RG shell ejections.
The  periodicity of the maxima is  in fact of  $\sim$ 300-400 days.

 During the 2000-2002  outburst of Z And,
  broad wings (up to 2000 \kms) developed in  the \Ha line profile (Tomov et al 2008). Moreover,
collimated bipolar jets appeared and disappeared  throughout the 2006 outburst (Skopal et al 2009).
Broad lines were   observed  in other SS, e.g. CH Cyg, BI Cru, AG Dra, etc. They 
originate from the high velocity matter  accompanying the outbursts.
Exceptionally broad \Ly and \Ha line profiles were explained by 
WD  explosion (Contini et al. 2009 a,c), while
 collimated jets at some epochs reveal the presence of the accretion disk. 
Both broad lines and jets were not observed during the 1984-1986 
outburst of Z And.

\section*{Acknowledgments}
We are  grateful to   Sharon Sadeh for helpful advise.

\end{document}